# DFT study of ionic liquids adsorption on circumcoronene shaped graphene

Anton Ruzanov,[1] Meeri Lembinen,[2] Heigo Ers,[1] José M. García de la Vega,[3] Isabel Lage-Estebanez,[3] Enn Lust[1] and Vladislav B. Ivaništšev[1]*

Address: [1] Institute of Chemistry, University of Tartu, Ravila 14a, Tartu 50411, Estonia, [2] Institute of Physics, University of Tartu, Ostwaldi str. 1, Tartu 50411, Estonia, [3] Departamento de Química Física Aplicada, Universidad Autónoma de Madrid, Ciudad Universitaria de Cantoblanco, 28049, Madrid, Spain.

Email: Vladislav B. Ivaništšev – vladislav.ivanistsev@ut.ee

* Corresponding author

## Abstract

Carbon materials have a range of properties such as high electrical conductivity, high specific surface area, and mechanical flexibility are relevant for electrochemical applications. Carbon materials are utilised in energy conversion-and-storage devices along with electrolytes of complementary properties. In this work, we study the interaction of highly concentrated electrolytes (ionic liquids) at a model carbon surface (circumcoronene) using density functional theory methods. Our results indicate the decisive role of the dispersion interactions that noticeably strengthen the circumcoronene–ion interaction. Also, we focus on the adsorption of halide anions as the electrolytes containing these ions are promising for practical use in supercapacitors and solar cells.

## Introduction

Understanding the interaction of ionic liquids (ILs) with carbon materials is crucial in the development of batteries and supercapacitors, $CO_2$ capture-and-storage devices, superconducting devices and electromechanical systems [1,2]. Non-periodic models containing ionic associates at polycyclic aromatic hydrocarbons (PAHs) are used to study the various surface phenomena taking place at the carbon | electrolyte interfaces [3–5]. The partial charge transfer process between the carbon materials and adsorbed ions are an important feature of these systems [5]. Although it can be estimated with the density functional theory (DFT), the predicted transfer degree depends on the calculation method. This is partially due to the self-interaction error (SIE) [6]. Only in a few studies of ILs, the SIE effect on the DFT calculations was examined [7–9]. Grimme *et al.* concluded that the SIE error is almost negligible for three studied ionic pairs [7]. For a larger set of ionic associates, Lage-Estebanez *et al.* showed that the SIE is pronounced in the case of specific combinations of ions, in particular for ionic pairs including halide anions or pyridinium cation [8,10]. Similar studies of a broad set of organic molecules and halide ions indicated that the SIE is notable [11–13]. Its effect was observable in terms of partial charge transfer between atoms/ions and PAHs [5]. In this work, we have evaluated the partial charge transfer at a model carbon | electrolyte model interface consisting of a PAH and ionic pairs.

PAHs represent a class of carbon materials that are intermediate between graphene and porous carbons. Currently, they are among the most studied organic semiconductors due to the appealing characteristics of their electronic structure [14,15]. A wide variety of materials containing aromatic hydrocarbon is used in batteries because carbon-containing structures exhibit both higher specific charges and more negative redox potentials than most of the metal oxides and chalcogenides [16]. Wherein, the partial charge transfer process between the adsorbing ions and the electrode surface may become a useful phenomenon affecting the electronic properties of the surface, especially in the case of superconductors [17].

The partial charge transfer from an adsorbate to a PAH depends on the size and structure of the PAH. In computations, it also depends on the methods used for calculations and analysis. Baker and Head-Gordon observed that DFT methods overestimate the transfer due to the SIE, whereas Hartree–Fock (HF) method underestimates transfer due to overlocalization [5]. Below we also demonstrate that in the case of ILs, containing halide anions, the DFT can overestimate the amount of the partial charge transfer degree affecting the calculated properties of these systems.

To investigate the effect of the SIE on the DFT calculations involving ILs, we have selected two sets of imidazolium-based ionic pairs: [EMIm$^+$][BF$_4^-$, PF$_6^-$, AlCl$_4^-$] and [EMIm$^+$][Cl$^-$, Br$^-$, I$^-$]. The latter set consists of ionic pairs that are strongly affected by the SIE [8]. The corresponding ILs are commonly used in experimental as well as computational studies [18–31]. For example, an electrolyte containing AlCl$_4^-$ anions was proposed for ultrafast rechargeable aluminium-ion battery [18,19]. Very high energy density and power density values were achieved for supercapacitors using EMImBF$_4$, LiPF$_6$ or NaPF$_6$ based electrolytes [20–22]. A noticeable increase in power density (~20–30%) and shorter charging-discharging times were demonstrated for supercapacitors based on mixed ILs [20]: $EMImBF_4$ + EMImI, $EMImBF_4$ + EMImBr, $EMImBF_4$ + EMImCl. These ILs are also demonstrating reversible specific adsorption with partial charge transfer effect at metal as well as at highly oriented pyrolytic graphite surfaces [23–27]. More recently, using *in situ* synchrotron radiation based XPS method, a very well pronounced interaction of $I^-$ with micro-mesoporous carbon surface has been demonstrated [4,28]. Future experiments with $Br^-$ or $Cl^-$ containing mixtures are important for both fundamental understandings as well as practical utilisation of carbon–electrolyte interfaces with characteristic partial charge transfer step. In complement to the previous experimental works, in this article, we present results of DFT calculations of six ionic pairs at the circumcoronene surface.

The article is organised as follows. In the Theoretical methods section, the computational details are provided. Effect of the SIE on adsorption and interaction energy and ionic charges for six circumcoronene–ionic pairs are presented in the Results and Discussion section. Comparison of these results with the available experimental and computational data along with a discussion of charge analysis finalises this section. Finally, the main outcomes of this work are summarised in the Conclusions section.

## Theoretical methods

The effect of the SIE was investigated on six ionic pairs at the circumcoronene ($C_{54}$). Such relatively small non-periodic systems are suitable for calculations with hybrid functionals that alleviate the SIE. The ionic pairs consisted of 1-ethyl-3-methylimidazolium = $EMIm^+$ cation and chloride = $Cl^-$, bromide = $Br^-$, iodine = $I^-$, tetrafluoroborate = $BF_4^-$, hexafluorophosphate = $PF_6^-$ or tetrachloroaluminate = $AlCl_4^-$ anions. The geometry of an ionic pair on top of fixed circumcoronene (as given in Figure 1) was optimised starting from a typical π-stacking distance of 3.5 Å between the imidazolium ring and the surface plane. All consequent single point calculations were performed based on optimised geometries.

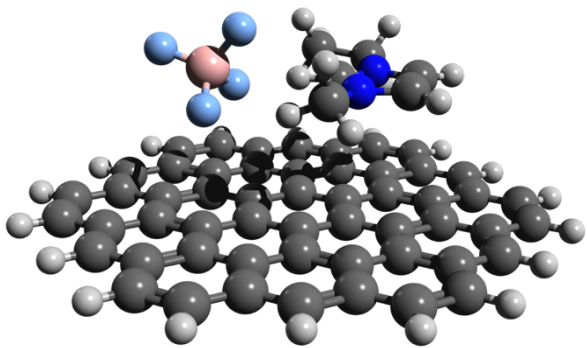

**Figure 1:** Optimised circumcoronene–[EMIm][$BF_4$] structure. The figure was prepared using Avogadro software [32].

All DFT calculations were run using the ORCA 4.0.0 program [33] using Perdew–Burke–Ernzerhof (PBE) [34] functional and its reparametrized version (PBEh-3c) [35]. The optimisation was run with the PBEh-3c method using double-ζ basis set (def2-mSVP), Grimme's dispersion correction (D3) [36,37], and geometrical counterpoise correction (gCP) [38]. Grimme *et al.* developed the PBEh-3c functional to fill a gap between existing semi-empirical methods and large basis set DFT calculations regarding cost-accuracy [35]. PBEh-3c does not suffer significantly from the SIE and represents an alternative to semi-local functionals in such problematic cases. Grimme *et al.* reported that by a systematic cancellation of errors between the density functional and the applied small Gaussian orbital basis set, accurate molecular structures, including bond lengths, can be obtained. In this work, we use

PBEh-3c functional as an alternative to range-separated hybrid functionals in suppressing the SIE [8].

The Charges from Electrostatic Potentials using Grid-based method (CHELPG) [39] was used for the estimation of charges in circumcoronene–ionic pair systems. Interaction energy $E_{int}$, adsorption energy $E_{ads}$ and dissociation energy $E_{diss}$ were calculated to describe the processes depicted in Figure 2.

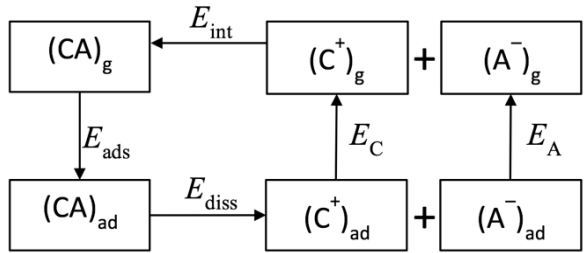

**Figure 2:** Processes occurring between ions of ionic liquid on the circumcoronene surface and away from it. The geometries of single ions and ionic pairs in vacuum were fixed and identical to the optimised geometries at the circumcoronene surface.

The interaction energies of ions in the solution $\Delta E_{int}$ were defined as the difference between the energy of the ionic pair (CA), cation (C), and anion (A):

$$\Delta E_{int} = E(\text{CA}) - E(\text{C}) - E(\text{A}) \tag{1}$$

The adsorption energy $\Delta E_{ads}$ was defined as the difference between the energies of the circumcoronene–ionic pair associate ($E(\text{C}_{54}\text{–CA})$), ionic pair (CA), and circumcoronene ($\text{C}_{54}$):

$$\Delta E_{ads} = E(\text{C}_{54}\text{–CA}) - E(\text{CA}) - E(\text{C}_{54}) \tag{2}$$

The dissociation energy of an IL associate on the circumcoronene surface $\Delta E_{diss}$ was defined as the difference between the sum of the energies of circumcoronene ($\text{C}_{54}$) and the circumcoronene–ionic pair associate ($E(\text{C}_{54}\text{–CA})$), the circumcoronene–cation ($\text{C}_{54}\text{–C}$), and circumcoronene-anion ($\text{C}_{54}\text{–A}$):

$$\Delta E_{diss} = E(C_{54}\text{–}CA) + E(C_{54}) - E(C_{54}\text{–}C) - E(C_{54}\text{–}A) \quad (3)$$

According to the diagram in Fig. 2, the sum of desorption energies of individual anion and cation from the circumcoronene can be expressed through $\Delta E_{int}$, $\Delta E_{ads}$, $\Delta E_{diss}$ as:

$$E_A + E_C = \Delta E_{diss} - \Delta E_{ads} + \Delta E_{int} \quad (4)$$

## Results and Discussion

### Geometries

The computations show that the imidazolium ring orients parallel to the circumcoronene surface. The distance between the N atoms in EMIm$^+$ and the surface plane is 3.32 Å for all studied systems, which is similar to a graphite interlayer separation of ~3.35 Å. This hints to the π–π stacking interaction between the imidazolium ring and circumcoronene. The distance between the surface plane and halide anions increases in the row Cl$^-$ (3.11 Å) < Br$^-$ (3.36 Å) < I$^-$ (3.65 Å). In the molecular anions, fluoride and chlorine atoms that are closer to the surface plane are located at a distance of 2.82 Å for BF$_4^-$ and PF$_6^-$ anions and 3.26 Å for AlCl$_4^-$ anion. The C−H⋯F/Cl/Br/I hydrogen bond length in the ILs containing halide anions increases in the row Cl$^-$ (1.97 Å) < Br$^-$ (2.27 Å) < I$^-$ (2.55 Å) and for molecular anions in the row BF$_4^-$ (1.83 Å) < PF$_6^-$ (1.99 Å) < AlCl$_4^-$ (2.57 Å).

### Adsorption, interaction, and dissociation energies

Table 1 presents adsorption ($E_{ads}$), interaction ($E_{int}$), and dissociation ($E_{diss}$) energies calculated in this work. Table 2 summarises the data from Refs. [3,40–44]. Overall, for the same surface model, the adsorption energy for ILs containing halide and molecular anions is roughly the same. The big difference in the adsorption energies for Cl$^-$ and PF$_6^-$ containing ionic pairs was only reported by Ghatee and

Moosavi [40], where the halide anion undergoes some repulsion from the surface during the optimisation at the DFT/B3LYP/6-311g level of theory. This difference in the geometry explains the lower adsorption energy values obtained by Ghatee and Moosavi [40]. The weak interaction of ionic pairs with small PAHs observed in work [3] is also due to the different geometry of the studied adsorption complexes. In our calculations, all halide ions locate close to the surface and thus contribute more to the favourable interaction.

**Table 1:** Energies of ionic pair adsorption on circumcoronene ($E_{ads}$), anion–cation interaction ($E_{int}$) and ionic dissociation at the circumcoronene surface ($E_{diss}$). All values are calculated with the PBEh-3c functional and are given in kJ/mol. D3 denotes the Grimme's dispersion correction.

| Ionic pairs | $E_{ads}$ | | $E_{int}$ | | $E_{diss}$ | |
|---|---|---|---|---|---|---|
| | $E_{ads}$ | D3 | $E_{int}$ | D3 | $E_{diss}$ | D3 |
| EMImCl | −81.6 | −45.6 | −465.6 | −4.2 | 390.9 | −3.7 |
| EMImBr | −79.0 | −46.4 | −401.0 | −4.2 | 346.1 | −3.7 |
| EMImI | −79.2 | −47.3 | −363.3 | −4.6 | 321.1 | −4.1 |
| EMImBF$_4$ | −85.0 | −47.8 | −367.0 | −6.0 | 318.6 | −5.3 |
| EMImPF$_6$ | −96.4 | −53.4 | −344.0 | −7.6 | 301.7 | −6.8 |
| EMImAlCl$_4$ | −95.3 | −54.1 | −330.2 | −11.1 | 295.2 | −10.3 |

**Table 2:** Graphene-like surface–ionic pair adsorption energies (kJ/mol). Top rows show data without dispersion correction. Bottom rows include results with Grimme's dispersion (D3) correction.

| Surface model/density functional | [MMIm$^+$, EMIm$^+$, BMIm$^+$, BPy$^+$, BMPyr$^+$] [Cl$^-$, Br$^-$] | [MMIm$^+$, EMIm$^+$, BMIm$^+$, BPy$^+$, BMPyr$^+$, BtMA$^+$] [PF$_6^-$, BF$_4^-$] | [EMIm$^+$, BMIm$^+$, BPy$^+$, BMPyr$^+$, BtMA$^+$] [TfO$^-$, Tf$_2$N$^-$, DCA$^-$] |
|---|---|---|---|
| C$_{54}$/B3LYP [41] | −30.9…−28.0 | −32.8…−21.2 | −33.8…−28.9 |
| h-BN/M06-2X [42] | | −49.5…−39.0 | −30.8…−26.6 |

| C$_{54}$/M06-2X [43] | | −62.8…−40.2 | −60.2…−44.1 |
|---|---|---|---|
| C$_{54}$/B3LYP [40] | −10.5…−10.1 | −89.6…−89.0 | |
| C$_{24}$/B3LYP [3] | −15.4…−9.7 | −16.4…−3.9 | −18.3…4.8 |
| C$_{66}$/BLYP+D3 [44] | | | −100 |
| C$_{54}$/M06-2X + D3 [43] | | −135.7…−108.6 | −118.9…−97.2 |

C$_{24}$ – coronene shaped graphene (C$_{24}$H$_{12}$); C$_{66}$ – coronene shaped graphene, (C$_{66}$H$_{20}$); h-BN – hexagonal Boron-Nitride, and the rest (MMIm, …DCA); MMIm – 1,3-dimethylimidazolium; BMIm – 1-butyl-3-methylimidazolium; BPy – 1-butylpyridinium; BMPyr – 1-butyl-1-methylpyrrolidinium; BtMA – butyl-trimethylammonium; TfO – trifluoromethane sulfonate; Tf$_2$N – bis(trifluoromethylsulfonyl)imide; DCA – dicyanamide.

The results obtained by other authors show that the dispersion correction makes a significant contribution to the adsorption energy. In fact, the values reported are much higher than those found in the present research. For example, additional results for choline benzoate ionic pair (not listed in Table 2) show that without the dispersion correction the interaction energy varies from −57.9 to −50.2 kJ/mol. However, these values with the dispersion correction, range from −166.9 to −146.6 eV [45]. Note that in works [3,40–44] BLYP, B3LYP, and M06-2X functionals were used. Nevertheless, we applied PBE and PBEh-3c functionals. BLYP is a GGA functional that is similar to PBE, while hybrid B3LYP and M06-2X functionals are similar to PBEh-3c.

In this work, $E_{ads}$ decreases in a row: EMImPF$_6$ ≈ EMImAlCl$_4$ > EMImBF$_4$ > EMImCl ≈ EMImBr ≈ EMImI, and the dispersion contributes to one-half of the adsorption energy (Table 1). For $E_{int}$ and $E_{diss}$ the situation is opposite – in the absence of the surface, the contribution of dispersion forces is relatively low. As expected, comparison of $E_{int}$ and $E_{diss}$ shows that the dissociation of an ionic pair at the circumcoronene requires less energy than in vacuum.

According to Eq. 4, the sum of $E_{ads}$, $E_{int}$, and $E_{diss}$ equals to the desorption energies of individual anion and cation ($E_A$ and $E_C$). $E_A$ and $E_C$ values were calculated for the optimised associate geometry by removing the counter-ion and assigning the charge. Apparently, the cation interacts with the surface much stronger than any of the anions. $E_C \approx 119$ kJ/mol for all associates. The desorption energy of anions (in kJ/mol) decreases in a row: $Cl^-$ (37.8) > $PF_6^-$ (19.8) ~ $Br^-$ (16.1) ~ $BF_4^-$ (14.8) ~ $AlCl_4^-$ (10.2) > $I^-$ (3.4). The row presents an interesting contradiction: from experiments, the iodide is known as a strongly adsorbing anion.

## Specific adsorption

The strong specific adsorption of halide ions at single crystal electrodes from different solvents was reported in many experiments [24,46–50]. The specific adsorption from ILs on carbon materials plays an important role in various electrochemical phenomena and produces a significant effect on the equilibrium parameters of the interphase boundary and the rate of electrochemical processes [51–53]. Recently, much attention has been dedicated to adsorption of halide anions because the presence of specifically adsorbing halide anions leads to the increase of interfacial capacitance and, therefore, the energy density of supercapacitors [24,46,47].

To understand what may lead to specific adsorption, we have computationally studied the adsorption of a single anion on neutral ($C_{54} + A^- \rightarrow C_{54}$–$A^-$) and charged ($C_{54}^+ + A^- \rightarrow C_{54}^{\delta+}$–$A^{\delta-}$) circumcoronene. Calculations were run after initial geometries optimisation by placing each ion on top of the centre of fixed circumcoronene. Table 3 shows the energies and charges for these two cases. In general, at the neutral surface, bromide and iodide show much lower tendency to adsorb (lower absolute $E_A$ values). On the contrary, in experiments, specific adsorption of these anions from ILs was observed [24,47,54,55]. It is worth mentioning that in calculations the minus charge of the $C_{54}$–$A^-$ associate is localised on the anion. The situation changes when the surface is polarised. In $C_{54}^{\delta+}$–$A^{\delta-}$ associate, the absolute charge of halide anions decreases almost to zero. Either halide atoms cannot accept the charge from the circumcoronene or the halide anions transfer their charge to the surface. In both interpretations, the result is the same, and it is in agreement with the experimental data, where at anodic polarisation halide anions not only specifically adsorbed but also oxidised [43]. It is worth noting, how the $E_A$ changes when one electron is removed from the system. For the molecular anions, the adsorption becomes more favourable due to the Coulomb attraction between the charged surface and the adsorbed anions. In the case of halide anions, the adsorption becomes even more favourable due to the circumcoronene–halide covalent bonding. Interestingly, at positive surface anions adsorb more strongly than EMIm$^+$ cation at the negative surface, although with neutral surface cation interacts much stronger than any of the anions.

**Table 3:** Energies of ionic pair adsorption on circumcoronene ($E_{ads}$), anion–cation interaction ($E_{int}$) and ionic dissociation at the circumcoronene surface ($E_{diss}$). All values are calculated with the PBEh-3c functional and are given in kJ/mol. D3 denotes the Grimme's dispersion correction.

| Anion | $C_{54}$–A$^-$ | | $C_{54}^+$–A$^-$ | |
|---|---|---|---|---|
| | $E_A$ | $q_A$ | $E_A$ | $q_A$ |
| Cl$^-$ | −40.9 | −0.75 | −377 | −0.03 |
| Br$^-$ | −16.9 | −0.81 | −345 | −0.10 |
| I$^-$ | −3.25 | −0.84 | −331 | −0.03 |
| BF$_4^-$ | −30.8 | −0.76 | −256 | −0.73 |
| PF$_6^-$ | −26.2 | −0.78 | −242 | −0.75 |
| AlCl$_4^-$ | −24.0 | −0.82 | −234 | −0.79 |

| Cation | $C_{54}$–C$^+$ | | $C_{54}^-$–C$^+$ | |
|---|---|---|---|---|
| | $E_C$ | $q_C$ | $E_C$ | $q_C$ |
| EMIm$^+$ | −122 | 0.77 | −205 | 0.74 |

Table 4 shows the total dipole moment of the studied systems in the direction perpendicular to the circumcoronene surface plane. The dipole originates from the specific arrangement of anions and cations on the charged surface. The dipole is also affected by the partial charge transfer between an ionic pair and circumcoronene. As expected, different DFT methods give a distinct estimation of the partial charge transfer. Both PBE and PBEh-3c show the similar trend. In the context of the interfacial potential drop ($\Delta U$), the latter is proportional to the interfacial dipole moment ($\mu$) as $\Delta U \sim \mu$ (discussed in detail in [56]). Different types of ions contribute to the interfacial dipole, obviously, to a variable extent. We may speculate that at a given surface charge the adsorption of halide anions from $EMImBF_4$ + EMImX mixtures would decrease the potential drop relative to its value in the pure $EMImBF_4$ ionic liquid, as in experiments [23,25,46]. Thus, the capacitance in these systems would increase for halide anions in a row: $I^- < Br^- < Cl^-$. Such simple estimation is in line with the experimental findings [48,49,57].

**Table 4:** Dipole moment values (in Debye).

|  | PBE | PBEh-3c |
|---|---|---|
| $C_{54}$–EMImCl | −0.06 | −0.39 |
| $C_{54}$–EMImBr | +0.23 | +0.01 |
| $C_{54}$–EMImI | +0.57 | +0.56 |
| $C_{54}$–$EMImBF_4$ | −0.80 | −0.53 |
| $C_{54}$–$EMImPF_6$ | −1.08 | −0.78 |
| $C_{54}$–$EMImAlCl_4$ | −2.22 | −2.11 |

**Charge distribution analysis**

Within the GGA, an ionic pair CA may dissociate into spurious fractional charge fragments $C^{\delta+}$ and $A^{\delta-}$, with an energy that is lower than of $C^0$ and $A^0$, if the lowest unoccupied orbital energy of anion (A) lies below the highest occupied orbital energy of the cation (C). The spurious fractional charge dissociation arises from the SIE – the failure of being exact for all one-electron densities is inherent for all the semilocal functionals. It was shown that this error is common for many ionic associates [58]. The SIE manifests itself most vividly in the long-range region of the interaction energy curve for diatomic associates as well as even at optimal distances for large PAH–alkali metal associates [5,59]. There are different ways to suppress the SIE, including utilisation of hybrid functionals, such as B3LYP, M06-2X, and PBEh-3c. Many of the previous theoretical results calculated for PAH complexes with IL ions employed pure GGA functionals which are sensitive to the SIE [3,5,40]. Therefore, the degree of the partial charge transfer in those works was probably overestimated. Calculations with pure PBE and hybrid PBEh-3c functional on the circumcoronene–ionic pair models provide a general outline of the problem.

The difference between using PBE and PBEh-3c functionals can be seen in the absolute charge values (Table 5). The charges obtained with the PBEh-3c functional are higher (in absolute scale) than those obtained with the PBE functional. The absolute partial charges show two sets of anions: halide anions with smaller charges, and anions with multiple atoms ($BF_4^-$, $PF_6^-$, $AlCl_4^-$) with bigger charges. The PBEh-3c functional, where SIE is taken into account compared to PBE, decreases the partial charge transfer which is most pronounced for the halide anions. Previously we established similar results showing that the SIE most strongly influences the calculation data of ionic associates that contain halide anions [8,60].

**Table 5:** Absolute charge of the anion (calculated with ChelpG, values in e).

|  | PBE | | PBEh-3c | |
| --- | --- | --- | --- | --- |
|  | $q_{CA}$ | $q_{C54-CA}$ | $q_{CA}$ | $q_{C54-CA}$ |
| $C_{54}$–EMImCl | −0.70 | −0.63 | −0.73 | −0.66 |
| $C_{54}$–EMImBr | −0.71 | −0.65 | −0.76 | −0.71 |
| $C_{54}$–EMImI | −0.70 | −0.64 | −0.78 | −0.75 |
| C$_{54}$–EMImBF$_4$ | −0.87 | −0.75 | −0.87 | −0.78 |
| C$_{54}$–EMImPF$_6$ | −0.89 | −0.78 | −0.89 | −0.80 |
| C$_{54}$–EMImAlCl$_4$ | −0.79 | −0.75 | −0.84 | −0.82 |

According to the literature, the charge of circumcoronene or similar surfaces with NBO/NPA ranges between −0.003 and 0.009 e [40,41], whereas that value with ChelpG varies between 0.01 and 0.088 e with imidazolium cations [3,43]. The charge differences ($\Delta q$) for BF$_4^-$ and PF$_6^-$ anions on graphene are −0.154 and −0.047 e, respectively [43] – those values are −0.026 and −0.029 e for C$_{24}$ and −0.035 e for Br$^-$ [3]. For all systems, the change for the BMIm$^+$ cation is between 0.041 and 0.145 e [3,43]. Shakourian–Fard *et al.* [43] observed that one of the most significant changes in ILs upon adsorption on the graphene surface is the change in the hydrogen-bond strength between the cation and the anion. They compared the charge of cation and anion in the IL before and after adsorption on the circumcoronene surface and found that the cation tends to gain charge (becoming less positive) and the anion to lose charge (being less negative) after adsorption on the circumcoronene surface.

Table 6 shows the electronic charge redistribution found in this work. The biggest charge distribution occurs with AlCl$_4^-$ anion-containing ionic pair on the circumcoronene surface, while for others the surface charge distribution is unnoticeable. Utilisation of the PBEh-3c functional decreases the partial charge transfer between anion and cation. Our results show smaller changes for the cations and less variation in the charge difference for anions than those reported in [43]. Note that the difference between PBE and PBEh-3c for the $\Delta q_{C54}$ is within the limits of computational error. In our opinion, the charge transfer between the circumcoronene and ionic pairs is insignificant. However, the inter-ionic charge redistribution is pronounced and is sensitive to the DFT method used. Here is worth to note, that results were calculated on identical geometries to those obtained by optimisation of ionic pairs on the circumcoronene surface.

**Table 6:** Absolute partial charge of the anion (calculated with ChelpG, values in e).

|  | PBE | | | PBEh-3c | | |
| --- | --- | --- | --- | --- | --- | --- |
|  | $\Delta q_{C54}$ | $\Delta q_A$ | $\Delta q_C$ | $\Delta q_{C54}$ | $\Delta q_A$ | $\Delta q_C$ |
| $C_{54}$–EMImCl | −0.006 | −0.066 | 0.073 | 0.012 | −0.068 | 0.056 |
| $C_{54}$–EMImBr | −0.013 | −0.061 | 0.074 | 0.002 | −0.051 | 0.049 |
| $C_{54}$–EMImI | −0.017 | −0.057 | 0.074 | −0.013 | −0.037 | 0.049 |
| $C_{54}$–EMImBF$_4$ | −0.020 | −0.117 | 0.137 | 0.003 | −0.089 | 0.086 |
| $C_{54}$–EMImPF$_6$ | −0.024 | −0.112 | 0.136 | 0.007 | −0.093 | 0.086 |
| $C_{54}$–EMImAlCl$_4$ | −0.043 | −0.046 | 0.088 | −0.031 | −0.022 | 0.054 |

## On the molecular dynamics simulations of carbon–electrolyte interfaces

In this work, we have studied a carbon–electrolyte interface using a rather sophisticated DFT method (PBEh-3c), yet a rather simple model (circumcoronene–ionic pair). Another popular computational method – molecular dynamics (MD) simulations – allows for studying much more complex carbon–electrolyte models, yet neglecting the electronic effects. The MD methods are actively used to provide molecular-level insights, for example into the 1) reorganization at the interface [61–63], 2) charging in pores [64–66], 3) influence of additives (water) [67,68], 4) adsorption [31,69,70].

There are constant discussions on the applicability of the MD method for describing realistic interfaces. Firstly, justification of the level of coarse-graining is needed [71]. Secondly, the justification for neglecting the electronic effects is required, in particular, polarizability in the ionic liquid bulk, in the surface layers and at the interface [72,73]. Our study provides some ground for such discussions.

1) The circumcoronene–ionic pair interaction leads only to a very minimal change in the circumcoronene atomic charge values. The estimated partial charge transfer values are different for PBE and hybrid PBEh-3c functionals, however, the absolute values are within the error estimations. Thus the electronic polarization might be neglected at neutral interfaces, even for halide anions. However, the polarization between anions and cations changes in the vicinity of the surface requires special attention.

2) The covalent bond formation is observed between all studied halide anions and charged circumcoronene surface. Therefore, in case of MD simulations, a reactive force-field is needed to capture the covalent bond formation, *i.e.* the specific adsorption. To our best knowledge, there were very few studies of ILs with such force-fields [74].

## Conclusion

In this work, we have investigated the adsorption behaviour of six EMIm$^+$[Cl$^-$, Br$^-$, I$^-$, BF$_4^-$, PF$_6^-$, AlCl$_4^-$] ionic pairs on the circumcoronene using novel PBEh-3c density functional. Our main findings are the following:

1) The dispersion makes a substantial contribution to the circumcoronene–ionic pair interaction – up to 50% at the DFT/PBEh-3c level of theory.

2) Single EMIm$^+$ cation interacts much stronger with the neutral circumcoronene than any of the anions. At a charged surface, the interaction of anions is greater than that of the EMIm$^+$ cation.

3) Molecular anions (BF$_4^-$, PF$_6^-$, AlCl$_4^-$) adsorb more strongly than the halide anions at the neutral circumcoronene. Upon surface charging, due to the Coulomb attraction, the circumcoronene–anion interaction becomes stronger.

We paid much attention to adsorption of halide anions because of its importance for both fundamental understanding as well as practical utilisation of carbon–electrolyte interfaces with partial charge transfer step. Our most important observation explains the specific adsorption of the halide anions at the graphitic surfaces: 5) *at a charged surface, for halide anions, the interaction becomes much stronger than for the molecular anions due to the circumcoronene–halide anion covalent bonding.*

The reported insight provides useful information on the ILs adsorption on the circumcoronene surface and their interaction mechanisms which may help to find links between the chemical structure of the carbon–electrolyte interfaces and their performance in energy conversion devices. Even though the real world is more complex and additional factors need to be considered, the present investigation shows a possible direction for future theoretical and experimental examinations.

## Acknowledgements

This work was supported by the EU through the European Regional Development Fund under project TK141 "Advanced materials and high-technology devices for energy recuperation systems" (2020.4.01.15-0011), by the Estonian Research Council (institutional research grant No. IUT20-13), by the Estonian Personal Research Project PUT1107 and by short-term scientific missions funded by COST actions MP1303 and CM1206. This work has been supported by Graduate School of Functional materials and technologies receiving funding from the European Regional Development Fund in the University of Tartu, Estonia.

Results were obtained in part using the High-Performance Computing Center of the University of Tartu, in part using the EPSRC funded ARCHIE-WeSt High-Performance Computer (www.archie-west.ac.uk, EPSRC grant no. EP/K000586/1), in part using Centro de Computacion Cientifica in Universidad Autonoma de Madrid (CCC-UAM). ILE and JMGV thank the "Comunidad de Madrid" (Project: LIQUORGAS-S2013/MAE-2800) for financial support.